# ADVANCING MEDICAL EDUCATION THROUGH THE CINNAMON WEB APPLICATION

**Iuliana Marin**

*National University of Science and Technology Politehnica Bucharest, Faculty of Engineering in Foreign Languages (ROMANIA), marin.iulliana25@gmail.com*

## Abstract

The cINnAMON EUREKA Traditional project endeavours to revolutionize indoor lighting positioning and monitoring through the integration of intelligent devices and advanced sensor technologies. This article presents the prototypes developed for various project components and explores their potential application in medical education, particularly for aspiring healthcare professionals. The current variant of the intelligent bulb prototype offers a comparative analysis of the project's bulb against commercially available smart bulbs, shedding light on its superior efficiency and capabilities. Furthermore, the initial smart bracelet prototype showcases its ability to collect and analyse data from an array of built-in sensors, empowering medical students to evaluate fragility levels based on accelerometer, gyroscope, orientation, and heart rate data. Leveraging trilateration and optimization algorithms, the intelligent location module enables precise monitoring of individuals' positions within a building, enhancing medical students' understanding of patient localization in healthcare settings. In addition, the recognition of human activity module harnesses data from the bracelet's sensors to classify different activities, providing medical students with invaluable insights into patients' daily routines and mobility patterns. The user's personal profile module facilitates seamless user registration and access to the comprehensive services offered by the cINnAMON system, empowering medical students to collect patient data for analysis and aiding doctors in making informed healthcare decisions. With the telemonitoring system, medical students can remotely monitor patients by configuring sensors in their homes, thus enabling a deeper understanding of remote patient management. The cINnAMON web application serves as a powerful tool for medical students, presenting patient data in both tabular and graphical formats, allowing for comprehensive analysis and visualization. Moreover, medical students can set personalized thresholds for monitoring conditions, receiving timely notifications and alerts when patients' health parameters deviate from normal ranges. Finally, the prototype testing phase ensures the fulfilment of functional requirements, meticulous defect identification, and proactive risk mitigation, reinforcing the reliability and efficacy of the cINnAMON EUREKA Traditional project. By integrating the cINnAMON web application into medical education, aspiring healthcare professionals can harness its multifaceted features to enhance their clinical skills, broaden their understanding of patient care, and embrace the potential of cutting-edge technology in the field of medicine.

## 1 INTRODUCTION

In the ever-evolving landscape of technology and healthcare, the need for innovative solutions that bridge the gap between traditional education and cutting-edge advancements has become increasingly pronounced. Modern healthcare education stands at a crossroads, where the traditional approaches that have served as foundations for generations are being challenged by the boundless possibilities offered by emerging technologies. The fusion of intelligent devices, sensor-driven solutions, and data analytics has the power to redefine how medical students acquire knowledge, hone their clinical skills, and prepare for the challenges of a rapidly evolving healthcare landscape.

In the pursuit of transforming medical education, various solutions have been introduced to harness the potential of technology. For example, smart lighting solutions have been employed to enhance the learning environment [1, 2]. These lighting systems offer more than simple illumination. They enable customization of lighting conditions, with features like adjustable colour temperature, brightness control, and scheduled lighting changes. Such innovations have been shown to improve concentration, reduce eye strain, and create conducive spaces for studying [3]. Smart bulbs typically rely on Bluetooth or Wi-Fi connectivity, allowing users to control lighting conditions through mobile apps, thereby promoting an immersive and focused learning experience [4, 5].

Wearable devices have also played a significant role in healthcare education. Smartwatches, fitness trackers, and health-monitoring wearables have become popular tools for tracking vital signs, physical activity, and collecting valuable data for educational purposes [6, 7]. Equipped with an array of sensors, these devices offer insights into a person's health and activity levels, fostering a better understanding of health-related concepts [8]. Their application in medical education has often been limited to personal health tracking rather than providing a comprehensive platform for medical education.

Beyond hardware devices, mobile applications and web-based platforms have become prevalent in delivering educational content, facilitating remote learning, and providing access to medical resources [9]. These platforms offer a wealth of knowledge, making educational materials readily available [10, 11]. However, their potential for hands-on, practical learning experiences remains limited. It is within this context that the cINnAMON EUREKA Traditional project distinguishes itself [12, 13]. This innovative venture goes beyond the conventional boundaries of smart lighting and wearable technology, offering an integrated ecosystem that encompasses lighting, data collection, and advanced analytics tailored specifically for medical education. Through a combination of intelligent bulbs, smart bracelets, location modules, activity recognition, user profiles, telemonitoring, and a dedicated web application, the project unites technology and education in a novel manner. The cINnAMON EUREKA Traditional project seeks to address this challenge by indoor lighting positioning and monitoring through the integration of intelligent devices and advanced sensor technologies. In the following chapters, we will explore the transformative potential of this project within the context of medical education, shedding light on its distinctive features while contrasting it with existing solutions that strive to achieve similar objectives.

In the following sections, we will delve into the details of each project component, highlighting its exceptional features and capabilities while critically examining it in contrast to existing solutions. Our aim is to showcase how the cINnAMON EUREKA Traditional project represents a significant leap forward in preparing the next generation of healthcare professionals for a future marked by innovation, precision, and compassionate patient care. By doing so, this project sets a new standard in medical education, empowering students with the skills and knowledge needed to excel in the ever-evolving field of healthcare.

## 2  METHODOLOGY

Traditional methodologies have long relied on didactic lectures, hands-on clinical experiences, and apprenticeship models to impart knowledge and skills to aspiring medical practitioners [14]. The review encapsulates the conventional structure of medical curricula, emphasizing the importance of foundational sciences, clinical rotations, and the progression from theory to practical application. In addition to methodologies and tools, the review explores the conventional approaches to medical training. Mentorship and clinical preceptorships, for instance, have been instrumental in guiding medical students through the complexities of patient care [15, 16]. These approaches emphasize the importance of experiential learning, as students apprentice under experienced practitioners to develop clinical skills and cultivate a holistic understanding of medical practice.

By shedding light on these traditional facets of medical education, the innovative features of the cINnAMON EUREKA Traditional project can be discerned. It allows for an appreciation of the challenges faced by traditional educational models, such as limitations in real-time monitoring and dynamic engagement. This understanding becomes pivotal for recognizing how emerging technologies can bridge these gaps and enhance the educational journey for future healthcare professionals. The exploration of how technology has been integrated into medical education is a dynamic journey into the ever-evolving landscape of pedagogical innovation. The transformative impact of technology is evident across various domains, ranging from e-learning platforms to digital simulations, telemedicine applications, and the integration of medical devices within educational contexts [17, 18]. These platforms, often accessible online, offer a flexible and interactive approach to learning, enabling students to engage with educational content at their own pace. E-learning resources include multimedia presentations, virtual lectures, and interactive modules that cater to diverse learning styles, fostering a more inclusive and personalized educational experience.

The integration of telemedicine within medical education is a noteworthy development. This enables students to engage with patient cases, consultations, and medical discussions from diverse geographical locations [19]. This not only broadens the scope of medical education but also exposes students to a variety of patient cases and healthcare practices.

One of the critical aspects of the research is the identification of gaps and opportunities within the field of medical education. The recognition of these gaps not only serves as a foundation for understanding the evolving needs of medical education but also provides a contextual lens through which the cINnAMON EUREKA Traditional project can be evaluated in addressing these educational challenges. Emerging technologies offer opportunities for real-time feedback and assessment, addressing the challenge of monitoring patient progress [20]. The cINnAMON EUREKA Traditional project addresses the gap in real-time monitoring by integrating intelligent bulbs and smart bracelets. These components offer a novel way to track and assess patients' activities, allowing for immediate feedback and personalized insights about their health. Leveraging advanced sensor technologies in the smart bracelet, the project provides a platform for dynamic engagement through the collection and analysis of data from various sensors. This enables a more immersive learning experience, especially in areas such as patient monitoring and data analysis.

The cINnAMON web application serves as a powerful tool for comprehensive analysis and visualization of patient data. This addresses the limitations of traditional approaches by offering a centralized platform for medical students to explore, analyze, and understand the triggered data in both tabular and graphical formats. In summary, the literature review uncovers gaps in traditional medical education methods and highlights opportunities for improvement through technological integration [21]. These identified gaps provide a crucial context for assessing how the cINnAMON EUREKA Traditional project, with its innovative components and integrated technologies, contributes to addressing these educational challenges and fostering a more effective and engaging learning environment for aspiring healthcare professionals.

There are barriers and challenges that medical institutions and educators face when incorporating technology into their programs. These barriers include issues related to cost, infrastructure, resistance to change, and data security [22] that are not only respected in the medical domain, but also in other [23, 24]. Recognizing these challenges allows us to assess how the cINnAMON project addresses and mitigates similar issues. Our methodology ensures that we are well-informed about the existing landscape of medical education and the opportunities for innovation. The development of the intelligent location module incorporated trilateration and optimization algorithms to enable precise monitoring of individuals' positions within a building. The module's design aimed at enhancing medical students' understanding of patient localization in healthcare settings, ensuring accurate tracking without compromising privacy. The cINnAMON's module for the recognition of human activity was designed to classify different activities based on data from the smart bracelet's sensors. The development process involved creating algorithms for activity recognition to provide medical students with insights into patients' daily routines and mobility patterns, enhancing observational skills crucial in clinical settings.

Seamless user registration and access were prioritized in developing the personal profile module. This module ensures that medical students can easily access and utilize the comprehensive services offered by the cINnAMON system. The prototype development focused on user-friendly interfaces and efficient data management. The telemonitoring system was designed to allow medical students to remotely monitor patients by configuring sensors in their homes. The prototype development emphasized the integration of sensors, data transmission security, and user-friendly configurations for effective remote patient management.

## 3 RESULTS

The results section offers a comprehensive exploration of the cINnAMON EUREKA Traditional project, delving into its individual prototypes, and evaluating the integration of the web application into medical education settings.

### 3.1 Intelligent Bulb Prototype

The comparative analysis of the intelligent bulb prototype against commercially available smart bulbs unfolds a nuanced understanding of its capabilities [25]. Smart bulbs are built around a computing system based on the Raspberry Pi 3 model B (see Fig. 1). This device has the same capability as a regular bulb for managing the light intensity of its LEDs. It is also equipped with a sensor module to measure temperature, humidity, concentration of $CO_2$ and volatile organic compounds, ambient light, and dust level in the environment. In addition, it has a Passive Infrared Sensor (PIR) for motion detection and a PIR array sensor that can be used to determine the position of people inside the room. Like traditional bulbs, smart bulbs can also perform Bluetooth Low Energy (BLE) scans to detect nearby BLE devices. Finally, smart bulbs can communicate with central servers using Wi-Fi and Ethernet connections.

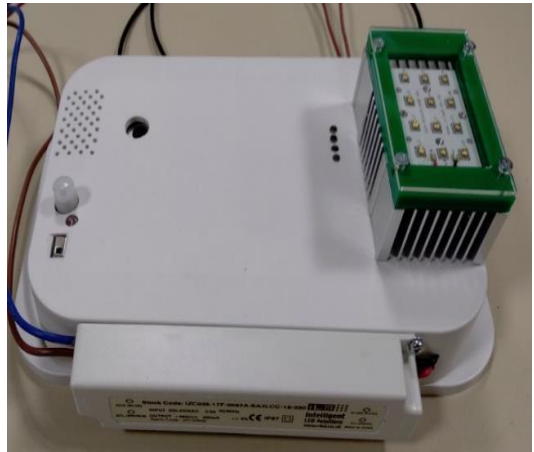

*Figure 1. cINnAMON smart bulb*

Our evaluation shows that direct trilateration is suitable for achieving room-level localization accuracy in realistic scenarios [26]. However, the raw readings showed significant variation. Studying the obtained distance estimates, we notice that in several situations a common intersection could not be identified, resulting in large error margins. This situation occurs both due to noisy RSSI readings caused by signal interference, as well as multipath fading and environmental factors. In the experiments undertaken, post-processing approaches improved the observed accuracy, by using the Kalman denoising filter. Compared to other indoor positioning systems, our proposed approach does not require the installation of additional devices or wiring and can achieve room-level accuracy in realistic scenarios that include signal pollution. Its implementation is simple and involves replacing some of the existing light bulbs with smart lighting fixtures.

## 3.2 Smart Bracelet Prototype

Empowering medical students to assess fragility levels based on a range of sensors, including accelerometer, gyroscope, orientation, and heart rate, the smart bracelet proved to be an asset for enhancing observational skills. The empirical testing phase validated the accuracy and reliability of the sensors, reinforcing the bracelet's role in providing valuable insights into patient mobility patterns. The companion application is a native JavaScript application developed in the Fitbit environment using the official API. Its main functions are to establish the communication link between the smartwatch (see Fig. 2), Fitbit cloud and distributed control system (DCS), to transmit the captured data for storage and processing; to act as the main node in the communications topology; open a two-way socket with Versa using native Fitbit methods, as well as a secure communication socket with DCS, via the NodeJS server; and to connect to the Fitbit cloud to retrieve relevant user information (name, age, weight, user ID), using OAuth 2 tokens for authentication.

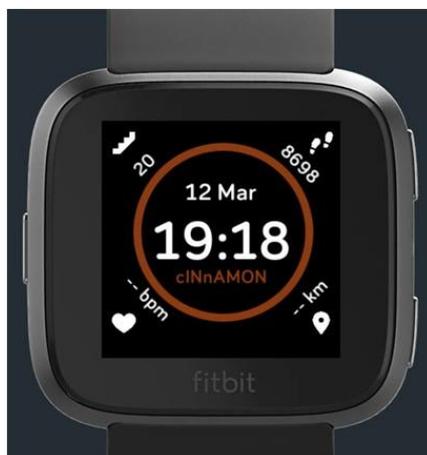

*Figure 2. cINnAMON smart bracelet screen*

## 3.3 Intelligent Location

The intelligent location module, leveraging trilateration and optimization algorithms, achieved precise monitoring of individuals' positions within a building (see Fig. 3). This module not only demonstrated technical robustness but also underscored its potential significance in medical education. The insights gained from monitoring individual positions enriched medical students' understanding of patient localization in healthcare settings, offering a novel perspective in educational scenarios.

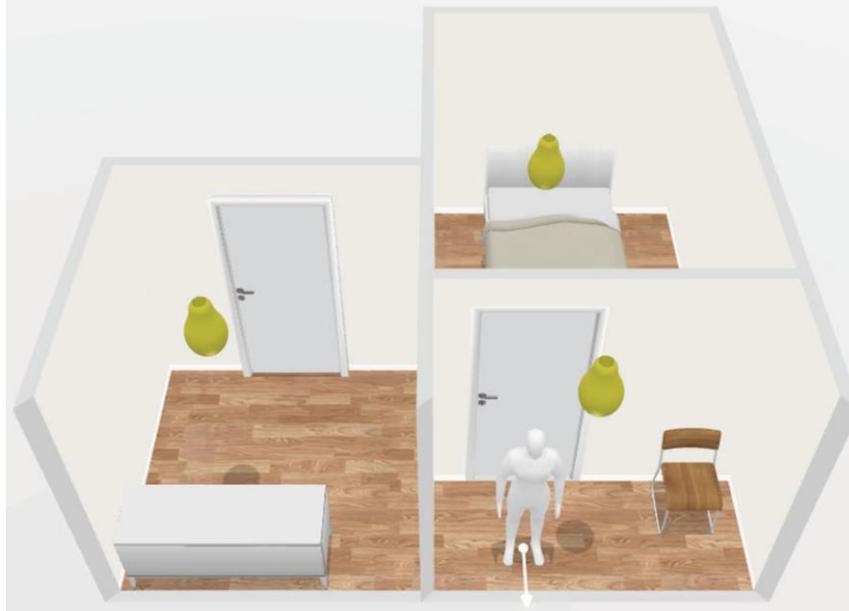

*Figure 3. Illustration of the indoor location of the person in the cINnAMON system interface*

Compared to existing systems, our solution integrates additional capabilities into a cost-effective device that can be easily deployed and looks unobtrusive in a typical home or office setting. In our experiments, the luminaires were mounted on the ceiling and indoor localization was attempted in three rooms using trilateration combined with filter-based approaches. Differences in room layouts, construction materials, the presence of furniture or large people, and signal interference from nearby devices cannot be fully accounted for outside of a laboratory setting.

## 3.4 Recognition of Human Activity

The recognition of human activity module emerged as a powerful tool for classifying different activities based on data from the smart bracelet's sensors. Beyond technical functionality, this module provided medical students with invaluable insights into patients' daily routines and mobility patterns. The empirical testing phase affirmed its effectiveness, positioning it as an asset in enhancing observational skills vital for aspiring healthcare professionals.

Data was collected for four types of activities: 1. Fast walking: the subject is walking at a fast pace; 2. Slow walking: the subject is walking at a slow pace; 3. Rest: the subject is still; 4. Climbing the stairs: the subject climbs a flight of stairs. Five one-minute recording sessions were conducted for these activities. A five-minute break was taken between sessions to avoid data compromise. The sensors (Fitbit Versa built-in sensors: accelerometer, gyroscope, orientation sensor, and optical sensors) were set to record data ten times per second (except for the optical sensor, which only recorded once per session).

The data was sent to a NodeJS server to be encapsulated in JSON format and saved to a MongoDB server for further processing. All data was recorded by the same user. The following classification algorithms were used for training: Logistic regression (LR); Decision Tree (DT); Random Forest (RF); Gradient Boosting (RF); K-nearest neighbors (KNN); Support Vector Classifier (SVC) and Gaussian Naive Bayes (GNB). Overall, Gradient Boosting outperformed the other models on all metrics, making it the best model for activity type recognition. Given that the data sets are time series, i.e., they are recorded one after the other, other models can be fitted to detect false positives and correct them during prediction.

## 3.5  User's Personal Profile

The seamless integration of the user's personal profile module facilitated efficient user registration and access to the cINnAMON system's comprehensive services. Beyond technical functionality, the user-friendly interface and robust data management features received positive feedback from medical students, contributing to an enhanced user experience. Once the administrator logs into the application, the list of users is presented in the form of a table, as illustrated in Fig. 4.

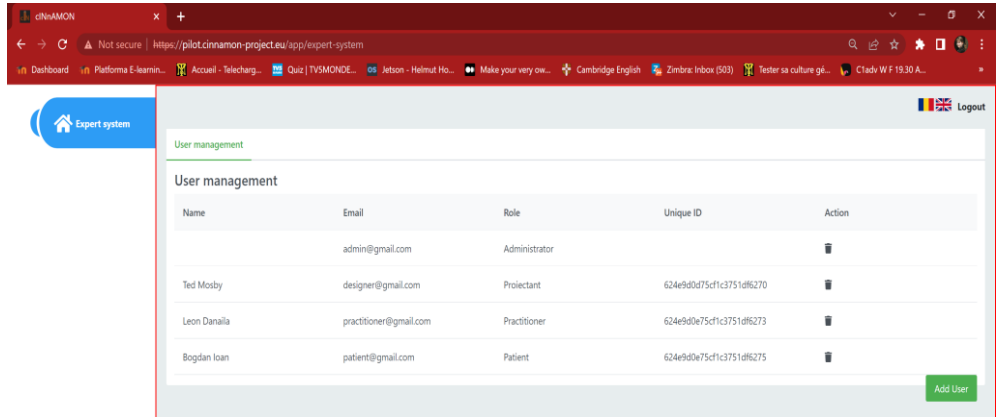

Figure 4. List of users

## 3.6  Telemonitoring

The telemonitoring system demonstrated its potential to reshape the landscape of remote patient management. Configuring sensors in patients' homes enabled medical students to remotely monitor health parameters, offering a deeper understanding of patient care beyond traditional clinical settings. This capability aligns with the evolving paradigm of healthcare delivery and underscores the project's practical relevance. A project is a digital representation of a patient's configuration, as illustrated in Fig. 5. To create this project, a designer must visit the patient's location, install the necessary sensors and accessories, and then enter them into the application. A single patient may have multiple locations, such as different rooms or floors, and may also have multiple sensors.

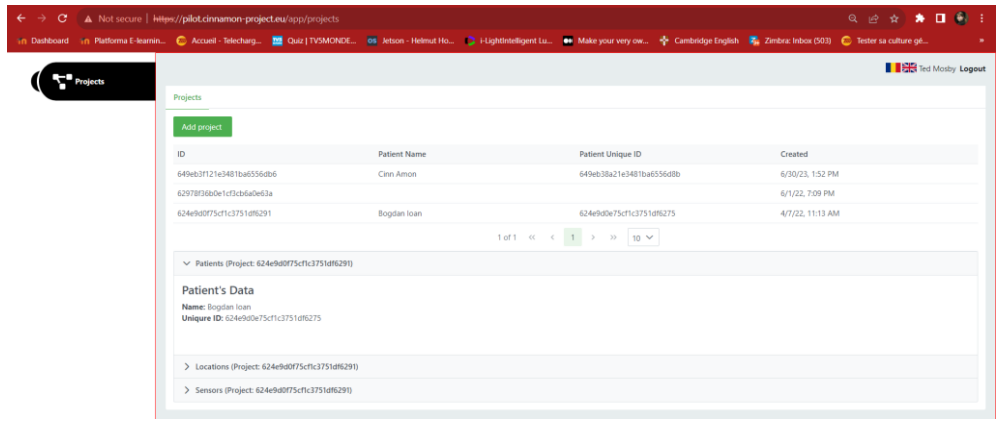

Figure 5. Patient home management

## 3.7  cINnAMON Web Application

The cINnAMON web application emerged as a powerful analytical tool. Presenting patient data in both tabular and graphical formats facilitated comprehensive analysis and visualization. Notably, medical students could set personalized thresholds for monitoring conditions, receiving timely notifications and alerts when patients' health parameters deviated from normal ranges, as in Fig. 6. This customization feature highlighted the adaptability of the web application to meet the diverse needs of healthcare learners.

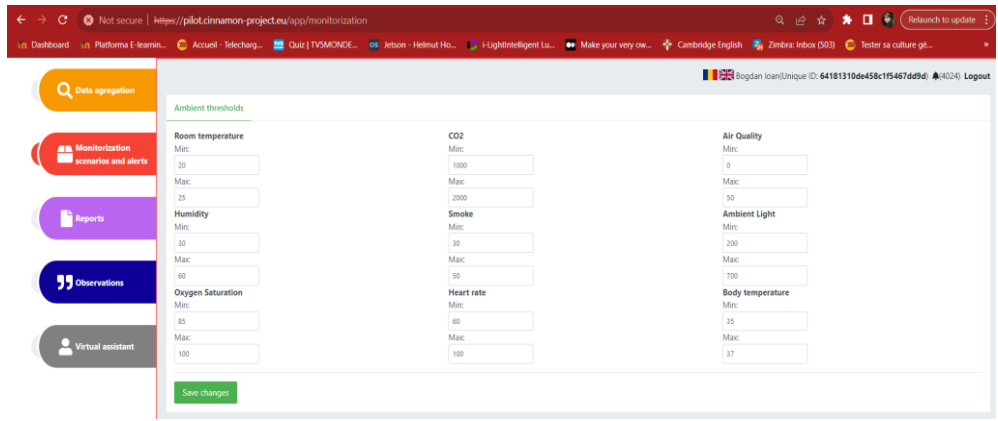

*Figure 6. Patient personalized thresholds*

The web application allows the doctor and medical student to analyse how the room temperature, CO2 level, air quality, humidity, smoke, and ambient light changed during a certain period.

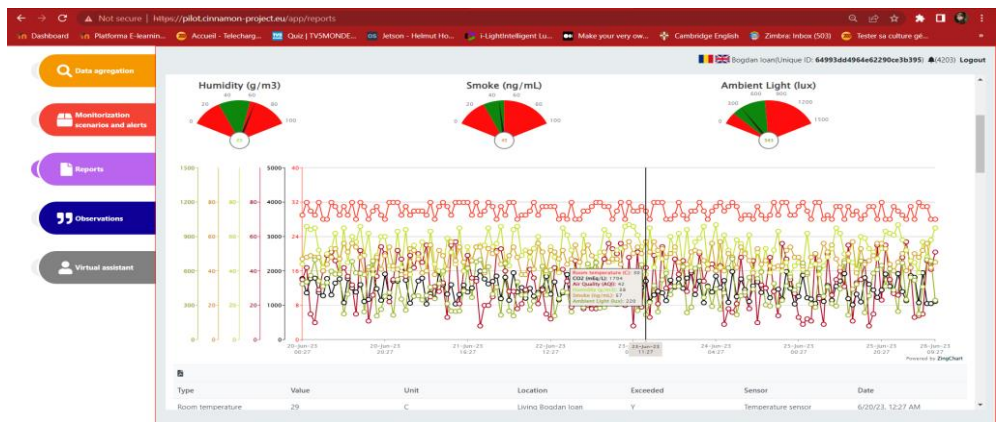

*Figure 7. Ambiental monitored parameters*

## 3.8 Prototype Testing

The prototype testing phase served as a stage in ensuring the fulfillment of functional requirements. Defect identification and proactive risk mitigation strategies reinforced the reliability and efficacy of the cINnAMON EUREKA Traditional project. This phase, critical for validating the project's readiness for practical implementation, is attested to its robustness and operational integrity. The test plan included defining the test strategy and managing the test process. To ensure exhaustive testing, test cases were developed according to the initial requirements. The test cases specified in the test plan were formulated in a generic way, without detailed specification of input and output data, making the plan easier to follow.

Test cases that failed were retested using the same input data as in the original test to verify that the error was reproducible. Also, these test cases were run multiple times, using various input data sets, to determine if the error is a general problem or caused only by some specific stimuli. On the user interface (UI) side, we focused our efforts on finding bugs in the interaction between the user and the application, especially at the graphical UI level.

Software testing is a method used to verify that the final software product meets the expected requirements and to ensure that it is free of defects. This process involves running software or system components using manual or automated tools to evaluate one or more key characteristics. The purpose of software testing is to identify bugs, gaps, or missing requirements compared to actual requirements.

In the pilot study, we used a Fitbit bracelet, with cINnAMON proprietary software, socio-demographic variables, and the Groningen fragility indicator [27]. The Groningen Indicator (GFI) is a validated questionnaire composed of 15 questions, with scores ranging from zero to fifteen. It assesses frailty in the physical, cognitive, social, and psychological domains. A GFI score of four or higher is considered a threshold for frailty.

Post-Study System Usability Questionnaire (PSSUQ) is a standardized questionnaire composed of 16 questions. It is widely used to assess users' perceived satisfaction with a website, software, system, or product at the end of a study. The questionnaire includes several sub-domains that provide useful information about the proposed system, including an overall assessment, system usefulness, information quality, and interface quality. A total of 16 participants took part in the study: 12 female participants and 4 male participants. From the analysis of the obtained scores, we can deduce that the study participants had a positive perception of their experience and the way the system works, both in general ("Overall" subscale) and in terms of the other three sub-domains: usefulness system (SYUSE), information quality (INFOQUAL) and interface quality (INTERQUAL).

## 4 CONCLUSIONS

In summary, the cINnAMON EUREKA Traditional project has emerged as a groundbreaking initiative in the convergence of indoor lighting technologies and medical education. The examination of individual prototypes, evaluations, testing, and integration into medical education settings has yielded insights into the transformative capacities of this undertaking.

The intelligent devices and sensor technologies, exemplified by the intelligent bulb and smart bracelet prototypes, signifies a pioneering solution for the dynamic requirements of medical education. These prototypes showcased performance, substantiating their potential to significantly enhance the educational journey of prospective healthcare professionals. Preliminary assessments of the cINnAMON system indicate a positive correlation between the integration of intelligent devices and the augmentation of clinical skills among medical students. The integration phase was marked by positive user feedback, emphasizing the intuitive interfaces and user-friendly attributes of the prototypes.

Future work will be dedicated to iterative improvements aimed at enhancing the performance, functionality, and overall user experience of the cINnAMON EUREKA Traditional project. The incorporation of user feedback and a proactive approach to technological advancements will inform ongoing development endeavours. Expanding beyond educational settings, future research will explore the potential clinical applications of the intelligent devices and sensor technologies. Investigating their integration into real-world healthcare scenarios aims to advance patient care and healthcare delivery. Ongoing development will focus on enhancing interactivity and personalization within the cINnAMON EUREKA Traditional project. Future iterations will explore integration possibilities with emerging technologies such as artificial intelligence, augmented reality, and virtual reality. This strategic integration is anticipated to further enrich the educational landscape for healthcare professionals.

In conclusion, the cINnAMON EUREKA Traditional project represents a transformative paradigm in medical education, orchestrating intelligent devices, sensor technologies, and a comprehensive web application. The affirmative outcomes underscore its potential to redefine the educational journey for aspiring healthcare professionals. Future work is poised to build upon these foundational achievements, striving for continuous improvement, expanded applications, and collaborative endeavours to further elevate the confluence of technology and medical education.

## ACKNOWLEDGEMENTS

The research for the current paper was funded by the Romanian Ministry of Education and Research, CCC DI-UEFISCDI, project number PN-III-P3-3.5-EUK-2019-0202, cINnAMON project, nr. 189/01.10.2020, within PNCDI III, as well as from the PubArt program of National University of Science and Technology Politehnica Bucharest, Romania.

## REFERENCES

[1]   B. Sun, Q. Zhang, S. Cao, "Development and Implementation of a Self-Optimizable Smart Lighting System Based on Learning Context in Classroom," *International Journal of Environmental Research and Public Health*, vol. 17, no. 4, pp. 1217, 2020.

[2]   A. G. Putrada, M. Abdurohman, D. Perdana, H. H. Nuha, "Machine Learning Methods in Smart Lighting Toward Achieving User Comfort: A Survey," *IEEE Access*, vol. 10, pp. 45137-45178, 2022.

[3]   L. Dębska, A. Białek, "Lighting Conditions as the Occupational Health Related Issue–Case Study," *MATEC Web of Conferences*, vol. 354, pp. 59, 2022.

[4] S. Cheruvu, A. Kumar, N. Smith, D. M. Wheeler, S. Cheruvu, A. Kumar, N. Smith, D. M. Wheeler, "Connectivity Technologies for IoT," *Demystifying Internet of Things Security: Successful IoT Device/Edge and Platform Security Deployment*, pp. 347-411, 2020.

[5] S. Ma, Q. Liu, P. C-Y. Sheu, "Foglight: Visible Light-Enabled Indoor Localization System for Low-Power IoT Devices," *IEEE Internet of Things Journal*, vol. 5, no. 1, pp. 175-185, 2017.

[6] C. V. Anikwe, H. F. Nweke, A. C. Ikegwu, C. A. Egwuonwu, F. U. Onu, U. R. Alo, Y. W. Teh, "Mobile and Wearable Sensors for Data-Driven Health Monitoring System: State-of-the-Art and Future Prospect," *Expert Systems with Applications*, vol. 202, pp. 117362, 2022.

[7] V. Vijayan, J. P. Connolly, J. Condell, N. McKelvey, P. Gardiner, "Review of Wearable Devices and Data Collection Considerations for Connected Health," *Sensors*, vol. 21, no. 16, pp. 5589, 2021.

[8] D. Dias, J. Paulo Silva Cunha, "Wearable Health Devices-Vital Sign Monitoring, Systems and Technologies," *Sensors*, vol. 18, no. 8, pp. 2414, 2018.

[9] I. Chatterjee, P. Chakraborty, "Use of Information Communication Technology by Medical Educators amid COVID-19 Pandemic and Beyond," *Journal of Educational Technology Systems*, vol .49, no. 3, pp. 310-324, 2021.

[10] J. C. Rawstorn, N. Gant, A. Meads, I. Warren, R. Maddison, "Remotely Delivered Exercise-Based Cardiac Rehabilitation: Design and Content Development of a Novel mHealth Platform," *JMIR mHealth and uHealth*, vol. 4, no. 2, pp. 5501, 2016.

[11] D. Guillaume, E. Troncoso, B. Duroseau, J. Bluestone, J. Fullerton, "Mobile-Social Learning for Continuing Professional Development in Low-and Middle-Income Countries: Integrative Review," *JMIR Medical Education*, vol. 8, no. 2, pp. 32614, 2022.

[12] B.-I. Ciubotaru, G.-V. Sasu, N. Goga, A. Vasilățeanu, A.-F. Popovici, "Architecture of a Non-Intrusive IoT System for Frailty Detection in Older People," *Electronics*, vol. 12, no. 9, pp. 2043, 2023.

[13] B.-I. Ciubotaru, G.-V. Sasu, N. Goga, A. Vasilățeanu, I. Marin, M. Goga, R. Popovici, G. Datta, "Prototype Results of an Internet of Things System Using Wearables and Artificial Intelligence for the Detection of Frailty in Elderly People," *Applied Sciences*, vol. 13, no. 15, pp. 8702, 2023.

[14] M. J. Hoskins, B. C. Nolan, K. L. Evans, B. Phillips, "Educating Health Professionals in Ultrasound Guided Peripheral Intravenous Cannulation: A Systematic Review of Teaching Methods, Competence Assessment, and Patient Outcomes," *Medicine*, vol. 102, no.16, 2023.

[15] A. L. Keinänen, R. Lähdesmäki, J. Juntunen, A. M. Tuomikoski, M. Kääriäinen, K. Mikkonen, "Effectiveness of Mentoring Education on Health Care Professionals' Mentoring Competence: A Systematic Review," *Nurse Education Today*, 2023.

[16] Q. Henry-Okafor, R. D. Chenault, R. B. Smith, "Addressing the Preceptor Gap in Nurse Practitioner Education," *The Journal for Nurse Practitioners*, 2023.

[17] G. Rouleau, M. P. Gagnon, J. Côté, J. Payne-Gagnon, E. Hudson, C. A. Dubois, J. Bouix-Picasso, "Effects of e-Learning in a Continuing Education Context on Nursing Care: Systematic Review of Systematic Qualitative, Quantitative, and Mixed-Studies Reviews," *Journal of Medical Internet Research*, vol. 21, no. 10, 2019.

[18] F. Ayoub, M. Moussa, A. G. Papatsoris, M. Abou Chakra, N. B. Chahine, Y. Fares, "The Online Learning in Medical Education: A Novel Challenge in the Era of COVID-19 Pandemic," *Hellenic Urology*, vol. 32, no. 2, pp. 89-96, 2020.

[19] A. M. Iancu, M. T. Kemp, W. Gribbin, D. R. Liesman, J. Nevarez, A. Pinsky, L. Pumiglia, J. J. Andino, H. B. Alam, J. N. Stojan, E. Walford, "Twelve Tips for the Integration of Medical Students into Telemedicine Visits," *Medical Teacher*, vol .43, no. 10, pp.1127-1133, 2021.

[20] J. Tyler, S. W. Choi, M. Tewari, "Real-Time, Personalized Medicine through Wearable Sensors and Dynamic Predictive Modeling: A New Paradigm for Clinical Medicine," *Current Opinion in Systems Biology*, vol. 20, pp. 17-25, 2020.

[21] K. T. Challa, A. Sayed, Y. Acharya, "Modern Techniques of Teaching and Learning in Medical Education: A Descriptive Literature Review," *MedEdPublish*, vol. 10, no. 18, 2021.

[22] M. Treve, "What COVID-19 has introduced into Education: Challenges Facing Higher Education Institutions (HEIs)," *Higher Education Pedagogies*, vol. 6, no. 1, pp. 212-227, 2021.

[23] M. Marin, L. Udres, E. Pogurschi, D. Dragotoiu, "Research Concerning the Influence of the Reducing Level of the Compound Feed on the Performances of the Pigs for Fattening," *Scientific Papers Animal Science and Biotechnologies*, vol. 43, no. 1, pp. 72-75, 2010.

[24] S. Demirkesen, A. Tezel, "Investigating Major Challenges for Industry 4.0 Adoption among Construction Companies," *Engineering, Construction and Architectural Management*, vol. 29, no. 3, pp. 1470-1503, 2022.

[25] I. Marin, A.-J. Molnar, "Evaluation of Indoor Localisation and Heart Rate Evolution," *International Conference on Computational Science and Its Applications*, pp. 75-89, 2021.

[26] I. Marin, M.-I. Bocicor, A.-J. Molnar, "Indoor Localization Techniques within a Home Monitoring Platform," *14th International Conference on Evaluation of Novel Approaches to Software Engineering*, pp. 378-401, 2020.

[27] I. Drubbel, N. Bleijenberg, G. Kranenburg, R. J. Eijkemans, M. J. Schuurmans, N. J. de Wit, M. E. Numans, "Identifying Frailty: Do the Frailty Index and Groningen Frailty Indicator Cover Different Clinical Perspectives? A Cross-Sectional Study," *BMC Family Practice*, vol. 14, pp. 1-8, 2013.